\documentclass[a4paper,12pt]{article}
\usepackage{amssymb}
\usepackage{epsf}
\usepackage{cite}

\makeatletter
\@addtoreset{equation}{section}
\makeatother

\def\preprint#1#2{\noindent\hbox{#1}\hfill\hbox{#2}\vskip 10pt}
\begin {document}
\begin{titlepage}

\preprint{ITP-UH-09/99}{April 1999}
\vfill

\begin{center}
{\Large\sc Doped Heisenberg chains: Spin-$S$ generalizations\\
  of the supersymmetric $t$--$J$ model}
\vfill

{\sc Holger Frahm}\footnote{e-mail: frahm@itp.uni-hannover.de}
\vspace{1.0em}

\emph{
Institut f\"ur Theoretische Physik, Universit\"at Hannover\\
  D-30167 Hannover, Germany}
\end{center}
\vfill

\begin{quote}
A family of exactly solvable models describing a spin $S$ Heisenberg
chain doped with mobile spin-(${S-1/2}$) carriers is constructed from
$gl(2|1)$-invariant solutions of the Yang-Baxter equation.  The models
are generalizations of the supersymmetric $t$--$J$ model which is
obtained for $S=1/2$.  We solve the model by means of the algebraic
Bethe Ansatz and present results for the zero temperature and
thermodynamic properties.  At low temperatures the models show spin
charge separation, i.e.\ contain contributions of a free bosonic
theory in the charge sector and an $SU(2)$-invariant theory describing
the magnetic excitations.  For small carrier concentration the latter
can be decomposed further into an $SU(2)$ level-$2S$
Wess-Zumino-Novikov-Witten model and the minimal unitary model ${\cal
M}_p$ with $p=2S+1$.
\end{quote}

PACS-Nos.\ {71.10.Pm, 
	75.10.Jm, 
	75.10.Lp, 
	75.30.Kz 
     }

Keywords: Spin-charge separation, Bethe Ansatz, $t$--$J$ model

\vspace*{\fill}
\setcounter{footnote}{0}
\end{titlepage}

%
\section{Introduction}
Doping of antiferromagnetic Mott insulators causes frustration which
has a profound effect on the magnetic properties of such systems.
High temperature superconductivity, charge ordering and anomalous
transport properties have been observed in doped transition metal
oxides with perovskite structure (see Ref.~\cite{dagx:99} for a
review).  Numerical studies of one-dimensional models proposed for
doped Nickel-oxides and Manganites for such systems showed strong
tendencies toward ferromagnetism and phase separation.  For a better
understanding of these phenomena requires to take into account strong
electronic correlations.
A commonly used starting point for a description of such systems is
the ferromagnetic Kondo Hamiltonian \cite{Zener51,AnHa55}
\begin{equation}
  {\cal H} = -t \sum_{\langle ij\rangle\sigma}
   \left(c^\dagger_{i\sigma} c_{j\sigma} + h.c.\right)
  - J_H \sum_i {{\sigma}}_i\cdot \mathbf{S}_i
\label{Hel}
\end{equation}
where the first term is the kinetic energy given in terms of canonical
fermionic creation and annihilation operators $c^\dagger_{i\sigma}$
and $c_{i\sigma}$ for the itinerant electrons on site $i$ with spin
$\sigma=\uparrow,\downarrow$ and the second is the ferromagnetic
Hund's rule coupling between the spins ${\sigma}_i^k =
\sum_{\alpha\beta} c^\dagger_{i\alpha} \left({\sigma}^k
\right)_{\alpha\beta} c_{i\beta}$, $k=x,y,z$, of the itinerant
electrons to a localized spin $\mathbf{S}_n$.  A Coulombic repulsion
to suppress double occupancy in the itinerant band is implicit.  Large
Hund coupling $J_H$ favours the alignment of the itinerant and the
localized spin, i.e.\ spin eigenstates with the maximum allowed total
spin.  Hence, for local spins of length $S-1/2$ the degrees of freedom
that need to be kept for an effective theory of the low lying modes in
this model are spin-$S$ ``spins'' and spin-$S-1/2$ ``holes''.
Within the double-exchange Hamiltonian \cite{Zener51,AnHa55} classical
``background spins'' $\mathbf{S}_i$ are used to approximate these
holes.  However, the non trivial phases arising in a full quantum
mechanical treatment of these spin degrees of freedom may induce
low-energy modes which are essential for an understanding of the
magnetic properties of these systems \cite{mhda:96}.

To obtain an effective lattice model on the $4S+1$ dimensional local
Hilbertspace one has to eliminate the other allowed spin
configurations in a perturbative analysis for $J_H\gg t$
\cite{dagx:96,mhda:96}.  To leading order in $J_H$ the effective
Hamiltonian of the resulting model is simply the projection of
(\ref{Hel}) onto the states listed above.  $SU(2)$-invariance implies
that this operator can be written as a polynomial of spin-operators:
\begin{equation}
  {\cal H}_{\rm eff} \approx -t \sum_{\langle ij\rangle}  
    {\cal P}_{ij} Q_S(y_{ij})\, \quad
  y_{ij} = \mathbf{S}_i\cdot\mathbf{S}_j/S(S-1/2)
\label{Heleff}
\end{equation}
where the operator ${\cal P}_{ij}$ permutes the states on sites $i$
and $j$ thereby allowing the holes to move.  For large but finite
$J_H$ one finds additional antiferromagnetic Heisenberg exchange
terms.  Hamiltonian operators of this type have been used as a
starting point for studies the phase diagram of doped transition metal
oxides by numerical diagonalization of small clusters
\cite{dagx:96,riera:97,dagx:99}.

In this paper we introduce a class of exactly solvable models in one
spatial dimension which generalize the supersymmetric $t$--$J$ model
\cite{lai:74,suth:75,schl:87} and a model for doped spin-$1$ chains
\cite{fpt:98}.  In spite of the appearance of several additional
couplings guaranteeing integrability these models may provide further
insights into the peculiar properties of these compounds.
In the following section we shall use the framework of the Quantum
Inverse Scattering Method (QISM) to construct families of vertex
models making use of so called `atypical' representations of the super
Lie algebra $gl(2|1)$.  The spectra of the corresponding commuting
transfer matrices are obtained by means of the algebraic Bethe Ansatz
(BA).  In Sect.~\ref{sec:fusion} we derive local Hamiltonians (i.e.\
operators involving nearest neighbour interactions on the lattice
only) similar to the ones discussed above using a fusion procedure.
In Sec.~\ref{sec:tba} integral equations determinining the spectra and
thermodynamic properties of these models are obtained from the BA
equations.  From these equations we obtain the phase diagram of the
doped spin chains in a magnetic field for low temperatures $T\ll H$
and the low temperature properties for vanishing field $H=0$ in
Sects.~\ref{sec:phasH} and \ref{sec:phasT}.  We conclude with some
remarks on a possible $SU(2)$-invariant effective field theory
description for the low energy/low temperature sector of these
systems.

\section{Construction of the models}
\label{sec:constr}
Below we will construct a class of integrable vertex models from
solutions to the Yang Baxter equation (YBE) which are invariant under the
action of the graded Lie algebra $gl(2|1)$ \cite{snr:77,marcu:80}.
The nine generators of $gl(2|1)$ are classified into even ($1$, $S^z$,
$S^\pm$, $B$) and odd ($V_\pm$, $W_\pm$) ones depending on their
parity w.r.t.\ grading.  The even generators are the spin operators
$S^\alpha$ form a $SU(2)$ subalgebra with commutation relations
$[S^z,S^{\pm}]\,=\,\pm S^{\pm}$, $[S^+,S^-]\,=\,2S^z$ and the $U(1)$
charge $B$ which commutes with the $S^\alpha$:
$[B,S^{\pm}]\,=\,[B,S^z]\,=\,0$.
The commutators between even and odd generators of the algebra are
\begin{eqnarray}
 \left[S^z,V_{\pm}\right]\,=\,\pm\frac{1}{2}V_{\pm},\quad
 \left[S^{\pm},V_{\pm}\right]\,=\,0,
&&
 \left[S^{\mp},V_{\pm}\right]\,=\,V_{\mp},\quad
 \left[B,V_{\pm}\right]\,=\,\frac{1}{2}V_{\pm},
\nonumber\\
 \left[S^z,W_{\pm}\right]\,=\,\pm\frac{1}{2}W_{\pm},\quad
 \left[S^{\pm},W_{\pm}\right]\,=\,0,
&&
 \left[S^{\mp},W_{\pm}\right]\,=\,W_{\mp},\quad
 \left[B,W_{\pm}\right]\,=\,-\frac{1}{2}W_{\pm},
\label{comm:eo}
\end{eqnarray}
and the odd generators satisfy \emph{anti}commutation rules
\begin{eqnarray}
  &&\{V_\pm,V_\pm\} = \{V_\pm,V_\mp\} 
        = \{W_\pm,W_\pm\} = \{W_\pm,W_\mp\}=0 \ ,
\nonumber\\
  &&\{V_\pm,W_\pm\} = \pm \frac{1}{2} S^\pm\ , \quad 
        \{V_\pm,W_\mp\} = \frac{1}{2}\left(-S^z \pm B\right) .
\label{comm:oo}
\end{eqnarray}
The 'typical' representations $[b,s]$ of this algebra can be
characterized by the eigenvalues of operators $B$ and $\mathbf{S}^2$
on the multiplet with largest total $SU(2)$-spin
\cite{snr:77,marcu:80}.  Their dimension is $8S$ and they can be
decomposed into two spin-$(S-{1/2})$ multiplets with charge $b\pm1/2$
and a spin-$S$ and a spin-$(S-1)$ multiplet with charge $b$ with
respect to the $SU(2)$-subalgebra of $gl(2|1)$.  In the following we
shall be particularly interested in the $(4S+1)$-dimensional so-called
'atypical' representations $[S]_+$ which contain two multiplets of
spin $S$ and $(S-{1/2})$ and corresponding charges $b=S$ and
$b=S+1/2$.  Choosing a basis $\{|b,s,m\rangle\}$ in which $B$,
$\mathbf{S}^2$ and $S^z$ are diagonal, the nonvanishing matrix
elements of the fermionic operators are
\begin{eqnarray}
  \langle S+{1\over2},S-{1\over2},m\pm{1\over2}| V_\pm |S,S,m\rangle
  &=& \pm \sqrt{\frac{S\mp m}{2}}
\nonumber\\
  \langle S,S,m| W_\pm |S+{1\over2},S-{1\over2},m\mp{1\over2}\rangle
  &=& \sqrt{\frac{S\pm m}{2}}\ .
\label{repVW}
\end{eqnarray}
Tensor products of atypical representation can be decomposed as
\begin{eqnarray}
  &&  \left[S\right]_+ \otimes \left[S'\right]_+ =
  \left[S+S'\right]_+ 
	\oplus \left[S+S'+{1\over2},S+S'-{1\over2}\right]
\nonumber\\
  &&\qquad
	\oplus \left[S+S'+{1\over2},S+S'-{3\over2}\right]
	\oplus\cdots
	\oplus \left[S+S'+{1\over2},|S-S'|+{1\over2}\right]\, .
\label{tensor}
\end{eqnarray}
The irreducible components in this tensor product can be identified by
the action of the quadratic Casimir of the algebra
\begin{equation}
  K_2 = {\mathbf S}^2
      - B^2 - W_-V_+ + W_+V_- - V_-W_+ + V_+W_-\,
\label{casimir2}
\end{equation}
which has eigenvalues 0 on $\left[s\right]_+$ and $(s^2-b^2)$ on
$\left[b,s\right]$.

Choosing an irreducible $d$-dimensional representation of $gl(2|1)$
acting on a quantum space ${\cal V}\sim \mathbb{C}^d$, it is
straightforward to verify that the ${\cal L}$-operator
\cite{kulish:85}
\begin{equation}
   {\cal L}(\mu) =  \left( \begin{array}{ccc}
           \mu+ 2iB	& i\sqrt{2}W_-	& i\sqrt{2} W_+ \\
	   i\sqrt{2}V_+	& \mu+i(B+S^z)	& -iS^+ \\
	  -i\sqrt{2}V_-	& -iS^-		& \mu+i(B-S^z)
	\end{array}\right)\ .
\label{lop}
\end{equation}
written as a matrix in the three-dimensional matrix space ${\cal M}$
solves the Yang-Baxter equation
\begin{equation}
  R(\lambda-\mu) \left({\cal L}(\lambda)\otimes{\cal L}(\mu)\right)
  =\left({\cal L}(\mu)\otimes{\cal L}(\lambda)\right) R(\lambda-\mu)
\label{ybe3S}
\end{equation}
with the $R$-matrix
\begin{equation}
  R(\lambda)=b(\lambda)I + a(\lambda) \Pi\ ,\quad
  a(\lambda)=\frac{\lambda}{\lambda+i}\ ,\quad
  b(\lambda)=\frac{i}{\lambda+i}\ .
\label{rmat33}
\end{equation}
Here $I$ and $\Pi$ are the unit and \emph{graded} permutation operator
acting on the tensor product ${\cal M}_1\otimes{\cal M}_2$ of matrix
spaces in which ${\cal L}$-operators act in (\ref{ybe3S}).  Assigning
Grassmann parities $\epsilon_i\in\{0,1\}$ to the basis of these spaces
the matrix elements of $\Pi$ are
\begin{equation}
  \Pi_{i_2,j_2}^{i_1,j_1} = (-1)^{\epsilon_{j_1}\epsilon_{j_2}} 
	\delta_{i_1j_2}\delta_{i_2j_1}\ .
\end{equation}
Similarly, the matrix elements of the operators acting on the tensor
product of these spaces pick up signs
\(  \left(A\otimes B\right)_{i_2j_2}^{i_1j_1} =
	(-1)^{\epsilon_{i_2}(\epsilon_{i_1}+\epsilon_{j_1})}
	A_{i_1j_1} B_{i_2j_2}
\)
due to the grading of the basis.
Considering the ${\cal L}$-operator as a linear operator acting on the
tensor-product of spaces ${\cal M}\otimes {\cal V}$ with the
fundamental three-dimensional representation $\left[1/2\right]_+$
acting on the matrix space its $gl(2|1)$-invariance can be established
by rewriting (\ref{lop}) as $\mu-iK_2$ in terms of the Casimir
operator (\ref{casimir2}) on the tensor product (up to a shift of the
spectral parameter).

The intertwining relation (\ref{ybe3S}) implies that the monodromy
matrix, defined as the matrix product
\begin{equation}
  {\cal T(\lambda)} = {\cal L}_L(\lambda){\cal L}_{L-1}(\lambda)
	\cdots{\cal L}_1(\lambda)
\label{mono33}
\end{equation}
of ${\cal L}_n$-operators (\ref{lop}) with entries acting on
different quantum spaces ${\cal V}_n$ satisfies a Yang-Baxter equation
with the same $R$-matrix (\ref{rmat33}):
\begin{equation}
  R(\lambda-\mu) \left({\cal T}(\lambda)\otimes{\cal T}(\mu)\right)
  =\left({\cal T}(\mu)\otimes{\cal T}(\lambda)\right) R(\lambda-\mu)\ .
\label{ybeT3S}
\end{equation}
As an immediate consequence of this identity the transfer matrix given
by the matrix super trace of ${\cal T}$
\begin{equation}
  t_{3}(\mu) = sTr\left({\cal T}(\mu)\right)
	=\sum_{i=1}^3 
	(-1)^{\epsilon_i} \left[ {\cal T}(\mu) \right]^{ii}
\label{trans3S}
\end{equation}
commutes for different values of the spectral parameter $\mu$ thus
being the generating functional for a family of commuting operators on
the graded tensor product of $L$ quantum spaces which we will identify
below with the Hilbert space of an integrable spin chain.  The
subscript to the transfer matrix is used to label the dimension of
the matrix space of the corresponding monodromy matrix. 

The spectrum of this transfer matrix is obtained by means of the
algebraic Bethe Ansatz (ABA) \cite{vladb}.  As a consequence of the
grading different sets of Bethe Ansatz equations (BAE) follow from
different orderings of the basis \cite{kulish:85}.  
We now restrict ourselves to representations $\left[S\right]_+$ in the
quantum spaces ${\cal V}_n$ where we choose the state
$|0\rangle_n\equiv|S,S,S\rangle_n$ as our reference state.  The two
other possible sets of BAE for this model are given in
Appendix~\ref{app:ba}, their equivalence is shown in
Appendix~\ref{app:eq}.

Since the ABA for the transfer matrix (\ref{trans3S}) is completely
analogeous to the case considered in \cite{kulish:85,esko:92,foka:93},
we only sketch the main steps leading to the BAE and the spectrum.
The action of (\ref{lop}) on the reference state is
\begin{equation}
  {\cal L}_n(\mu)|0\rangle_n = 
  \left(\begin{array}{ccc}
    \mu+2iS & 0 & 0 \\
       0 & \mu+2iS & 0 \\
     -i\sqrt{2}V_n^- & -iS_n^- & \mu
	\end{array}\right)|0\rangle_n\ .
\end{equation}
Similarly, acting with the monodrony matrix (\ref{mono33}) on the
state $|\Omega_S\rangle = |0\rangle_L\otimes\cdots\otimes|0\rangle_1$ we
get
\begin{equation}
  {\cal T}(\mu)|\Omega_S\rangle = 
  \left(\begin{array}{ccc}
    (\mu+2iS)^L & 0 & 0 \\
       0 & (\mu+2iS)^L & 0 \\
     C_1(\mu) & C_2(\mu) & \mu^L
	\end{array}\right)|\Omega_S\rangle\ .
\end{equation}
Hence, $|\Omega_S\rangle$ is an eigenstate of the transfer matrix
(\ref{trans3S}) with eigenvalue $(-\mu^L)$ (we have chosen the grading
$\epsilon_1=0$, $\epsilon_2=\epsilon_3=1$ in the matrix space here).
The operators $C_1(\lambda)$ and $C_2(\lambda)$ create a hole and
lower the spin of the system respectively.  For eigenstates of
$t_{3S}(\lambda)$ with $N_h$ holes (generating sites with spin
$S-{1/2}$) and magnetization $M^z=LS -{1\over2}N_h - N_\downarrow$ we
make the Ansatz
\begin{equation}
   |\tilde\lambda_1,\ldots,\tilde\lambda_n|F\rangle =
	C_{a_1}(\tilde\lambda_1)\cdots C_{a_n}(\tilde\lambda_n) 
	|\Omega_S\rangle F^{a_n\cdots a_1}
\label{BAvec}
\end{equation}
where $n=N_h+N_\downarrow$.  Using the algebra of the operators in
(\ref{ybeT3S}) we are led to an eigenvalue problem for the amplitudes
$F^{a_n\cdots a_1}$ which is solved by a second Bethe Ansatz
parametrized by 'hole rapidities'
$\{\tilde\nu_\alpha\}_{\alpha=1}^{N_h}$.  Finally, we find that
(\ref{BAvec}) is an eigenstate of (\ref{trans3S}) with eigenvalue
\begin{eqnarray}
  &&\Lambda_{3}\left(\mu|
	\{\tilde\lambda_j\}_{j=1}^{N_h+N_\downarrow},
	\{\tilde\nu_\alpha\}_{\alpha=1}^{N_h}\right) =
  -{\mu}^L \prod_{j=1}^{N_h+N_\downarrow}
	{\mu-\tilde\lambda_j+i \over \mu-\tilde\lambda_j}
\nonumber\\
 &&\qquad
 +\left({\mu+2iS}\right)^L\prod_{\alpha=1}^{N_h}
	{\mu-\tilde\nu_\alpha +i \over \mu-\tilde\nu_\alpha}
  \left\{ 1 - \prod_{j=1}^{N_h+N_\downarrow}
	{\tilde\lambda_j-\mu+i \over \tilde\lambda_j-\mu} \right\}
\end{eqnarray}
provided the spectral parameters $\tilde\lambda_j\equiv\lambda_j -iS$
and $\tilde\nu_\alpha\equiv\nu_\alpha-iS+i/2$ satisfy
the  following set of BAE
\begin{eqnarray}
   \left( {\lambda_j+iS\over \lambda_j-iS} \right)^L &=&
       \prod_{k\ne j}^{N_h+N_\downarrow}
	{{\lambda_j-\lambda_k+i}\over{\lambda_j-\lambda_k-i}}\,
       \prod_{\alpha=1}^{N_h}
	{{\lambda_j-\nu_\alpha-{i\over2}} \over
	 {\lambda_j-\nu_\alpha+{i\over2}} }\ ,
\nonumber\\
	&& \qquad j=1,\ldots,N_h+N_\downarrow
\label{bae3}\\
   1 &=& \prod_{k=1}^{N_h+N_\downarrow}
	{{\nu_\alpha - \lambda_k +{i\over2}} \over
	 {\nu_\alpha - \lambda_k -{i\over2}}}\ ,
   \quad \alpha=1,\ldots,N_h\ .
\nonumber
\end{eqnarray}

\section{Doped spin chains}
\label{sec:fusion}
Choosing the fundamental three dimensional representation
$\left[1/2\right]_+$ of $gl(2|1)$ in (\ref{lop}), the ${\cal
L}$-operator taken at $\mu=-i$ becomes a graded permutation operator
on the tensor product ${\cal M}\otimes{\cal V}$ of matrix and quantum
space.  Hence the transfer matrix (\ref{trans3S}) generates a
translation by one site on the lattice for this value of the spectral
parameter.  Expanding the logarithm of $t_3(\mu)$ about this shift
point we can construct a Hamiltonian with nearest neighbour
interactions only
(\ref{trans3S})
\begin{equation}
  -i\left.{\partial\over\partial\mu} \ln t_{3}(\mu)\right|_{\mu=-i}
  = \sum_{n=1}^L \ldots
\label{Hstj}
\end{equation}
which is the supersymmetric $t$--$J$ model (see e.g.\ \cite{esko:92}).
In this case, Eqs.~(\ref{bae3}) are known as Sutherland's form of the
BAE for this model \cite{suth:75}.

For representation different from $\left[{1/2}\right]_+$ it is not
possible to construct a local Hamiltonian directly from the ${\cal
L}$-operators (\ref{lop}) (they can be used to construct $t$--$J$
models perturbed by integrable impurities though
\cite{bef:96,sczv:97,schl:98}).  To obtain a homogenous lattice model
such as (\ref{Hstj}) new ${\cal L}$-operators have to be found which act on
tensor products of matrix and quantum spaces of the same dimension.
Their $gl(2|1)$-invariance implies that they can be expressed as sum
over multiples of the projectors on irreducible components of the
tensor product of representations in the two spaces.  Noting that the
\emph{spin} multiplets at charge $(S+S')$ in the tensor product
(\ref{tensor}) are just the ones obtained by adding two spins of
length $S$, $S'$ (and similarly spins $(S-1/2)$, $(S'-1/2)$ at charge
$(S+S'+1)$) we find linear operators ${\cal L}^{\{SS'\}} (\mu)$ acting
on spaces carrying atypical representations $\left[S\right]_+$ and
$\left[S'\right]_+$ from the $gl(2)$-invariant ones constructed in
Ref.~\cite{kulx:81}, namely:
\begin{equation}
   {\cal L}^{\{SS'\}}(\mu) =
	- \prod_{k=|S-S'|+1}^{S+S'} {{\mu-i k}\over{\mu+ik}}\,
	  {\cal P}_{\left[S+S'\right]_+}
	- \sum_{m=|S-S'|}^{S+S'-1} \,\,
	  \prod_{k=|S-S'|+1}^m {{\mu-i k}\over{\mu+ik}}\,
	  {\cal P}_{\left[S+S'+{1\over2},m+{1\over2}\right]}\, .
\label{lopss}
\end{equation}
Here ${\cal P}_\Lambda$ is a projector on the $gl(2|1)$-multiplet
$\Lambda$ in the tensor product $\left[S\right]_+\otimes
\left[S'\right]_+$.  Choosing one of the representations to be
$\left[1/2\right]_+$ and comparing this expression with (\ref{lop}) we
find 
\begin{equation}
  {\cal L}^{\{1/2,S\}}\left(\mu\right) =
	-{1\over\mu+i(S+1/2)}\, {\cal L}\left(\mu-i(S+1/2)\right)\, .
\end{equation}
The new ${\cal L}$-operators satisfy the YBEs
\begin{equation}
  R_{S_1S_2}(\lambda-\mu) 
	\left({\cal L}^{\{S_1S_3\}}(\lambda)
		\otimes{\cal L}^{\{S_2S_3\}}(\mu)\right)
= \left({\cal L}^{\{S_2S_3\}}(\mu) 
	\otimes{\cal L}^{\{S_1S_3\}}(\lambda)\right)
	R_{S_1S_2}(\lambda-\mu)
\label{ybeSSS}
\end{equation}
where $R_{SS'}=\Pi{\cal L}^{\{SS'\}}$.  As a consequence of this relation
the transfer matrices of the corresponding vertex models
\begin{equation}
  t^{\{S_0S\}}(\mu) 
	= sTr_0\left({\cal L}_L^{\{S_0S\}}(\mu) \cdots 
			{\cal L}_1^{\{S_0S\}}(\mu)\right)
\label{transSSS}
\end{equation}
(the product of ${\cal L}$-operators and the super trace are taken in
the $(4S_0+1)$-dimensional matrix space) commute, i.e. $\left[
t^{\{S_0S\}}(\mu), t^{\{S_1S\}}(\lambda) \right] =0$ for all $S_0$,
$S_1$.  From (\ref{lopss}) we observe that ${\cal L}^{SS}(\mu=0)
\propto \Pi$.  As in (\ref{Hstj}) an integrable Hamiltonian with
nearest neighbour interactions on the lattice can be constructed by
taking the logarithmic derivative of $\ln t^{\{SS\}}(\mu)$ at $\mu=0$.

The eigenstates of the transfer matrices $t^{\{S'S\}}(\mu)$ are
parametrized by the roots of the BAE (\ref{bae3}).  To compute their
eigenvalues we need the so-called fusion relations between these
operators for different $S'$ which are obtained from considering
tensor products of different matrix spaces.

Starting with the YBE (\ref{ybeSSS}) for $S_1=S_2=1/2$ we observe that
choosing $\lambda-\mu=-i$ the matrix $R_{{1\over2},{1\over2}}$ is
proportional to a projector onto the five-dimensional
subrepresentation $\left[1\right]_+$ of the tensor product
$\left[1/2\right]_+ \otimes \left[1/2\right]_+$.  This implies that
the ${\cal L}$-operator $\tilde{\cal L}(\mu)= {\cal L}^{\{1/2,S\}}
\left(\mu-i/2\right) \otimes {\cal L}^{\{1/2,S\}} \left(\mu+i/2\right)$
satisfies the condition
\begin{equation}
   {\cal P}_{\left[1\right]_+} \tilde{\cal L}(\mu) 
	{\cal P}_{\left[3/2,1/2\right]} \equiv 0\ .
\end{equation}
Consequently, it can be rewritten as
\begin{equation}
  \tilde{\cal L}(\mu) \sim
   \left(\begin{array}{cc}
	{\cal L}^{\{\left[3/2,1/2\right],S\}}(\mu) & * \\
	0 & {\cal L}^{\{1,S\}}(\mu)
   \end{array}\right)
\label{lopfus}
\end{equation}
after a proper reordering of the basis in ${\cal M}_1\otimes{\cal
M}_2$. Here ${\cal L}^{\{\left[3/2,1/2\right],S\}}(\mu)$ is a
$4\times4$ matrix acting on the $\left[3/2,1/2\right]$ component of
this tensor product.  Building a monodromy matrix from $L$ copies of
(\ref{lopfus}) we obtain the fusion relation for the corresponding
transfer matrices
\begin{eqnarray}
  \tilde{t}(\mu) &\equiv& sTr\left(
	\tilde{\cal L}_L(\mu) \ldots \tilde{\cal L}_1(\mu) \right)
\nonumber\\
	&=& t^{\{1/2,S\}}\left(\mu-{i/2}\right) 
	    t^{\{1/2,S\}}\left(\mu+{i/2}\right)
	= t^{\{1,S\}}(\mu) + t^{\{\left[3/2,1/2\right],S\}}(\mu)\,
\label{fuseq}
\end{eqnarray}
and an equivalent equation for their eigenvalues
$\Lambda^{\{\cdot\,S\}} (\mu)$.
Since there are still two unknown functions of $\mu$ on the RHS of
this equation it is not possible to determine the spectrum of the new
transfer matrices from (\ref{fuseq}) alone.  As additional information
we make use of the fact that the eigenvalues of the transfer matrix
are analytical functions of $\mu$, i.e.\ the residues at their simple
poles vanish as a consequence of the BAE (\ref{bae3}).  Complemented
by the trivial action of $t^{\{1,S\}}(\mu)$ and $t^{\{\left[3/2,
1/2\right], S\}}(\mu)$ on the pseudo vacuum $|\Omega_S\rangle$ this
allows to compute $\Lambda^{\{1,S\}}$ with the result (see e.g.\
\cite{pffr:96})
\begin{eqnarray}
  &&\Lambda^{\{1,S\}}\left(\mu|
	\{\lambda_j\}_{j=1}^{N_h+N_\downarrow},
	\{\nu_\alpha\}_{\alpha=1}^{N_h}\right) =
  \left( {\mu-iS\over\mu+iS}\,
	 {\mu-i(S+1)\over\mu+i(S+1)} \right)^L
	\prod_{j=1}^{N_h+N_\downarrow}
	{\mu-\lambda_j+i\over\mu-\lambda_j-i}
\nonumber\\
&&\quad
 -\left( {\mu-iS\over\mu+iS}\,
	 {\mu+i(S-1)\over\mu+i(S+1)} \right)^L
	\prod_{j=1}^{N_h+N_\downarrow}
	{\mu-\lambda_j+i\over\mu-\lambda_j}
	\prod_{\alpha=1}^{N_h}
	{\mu-\nu_\alpha-{i\over2}\over\mu-\nu_\alpha-{3i\over2}}
  \left\{1-\prod_{j=1}^{N_h+N_\downarrow}
	{\lambda_j-\mu+2i \over \lambda_j-\mu+i}
  \right\}
\nonumber\\
 &&\quad
 -\left( {\mu+i(S-1)\over\mu+i(S+1)} \right)^L
	\prod_{j=1}^{N_h+N_\downarrow}
	{\lambda_j-\mu+i\over\lambda_j-\mu}
	\prod_{\alpha=1}^{N_h}
	{\mu-\nu_\alpha+{i\over2} \over \mu-\nu_\alpha-{3i\over2}}
  \left\{1-\prod_{j=1}^{N_h+N_\downarrow}
	{\lambda_j-\mu+2i\over\lambda_j-\mu+i}
  \right\}\ .
\label{Lambda5}
\end{eqnarray}

As observed above, the local vertex operator ${\cal L}^{\{1,1\}}
(\mu=0)$ becomes a graded permutation operator on the tensor product
of the five-dimensional matrix space and the quantum space in which
the representation $\left[S=1\right]_+$ is acting.  Hence we can
proceed as for the the case of the fundamental representation above
and obtain the local lattice Hamiltonian for a spin-1 chain doped with
$S=1/2$ holes introduced in Ref.~\cite{fpt:98}:
\begin{equation}
   {\cal H}^{(1)} = -i \left.{\partial\over\partial\mu} 
	\ln\left(t^{\{1,1\}}(\mu)\right)\right|_{\mu=0} -3L
	=\sum_{n=1}^L \left\{
        {\cal H}^{\rm exch}_{n,n+1} 
	+ {\cal H}^{\rm	hopp}_{n,n+1}\right\}\,
	-N_h\, .
\label{hamil1}
\end{equation}
The exchange and kinetic part of the Hamiltonian expressed in terms of spin
operators ${\mathbf S_i}$ with ${\mathbf S}_i^2=S_i(S_i+1)$ with $S_i=1$
or ${1/2}$ read
\begin{eqnarray}
   {\cal H}^{\rm exch}_{ij} &=&
        {1\over2}\left(
         {1\over S_i S_j} {\mathbf S}_i \cdot {\mathbf S}_j - 1
         +\delta_{S_iS_j,1}\left(1-({\mathbf S}_i\cdot{\mathbf S}_j)^2\right)
        \right)\ ,
\nonumber\\
   {\cal H}^{\rm hopp}_{ij} &=& 
        -\left(1-\delta_{S_i,S_j}\right)
         {\cal P}_{ij} \left( {\mathbf S}_i \cdot {\mathbf S}_j \right)\ .
\nonumber
\end{eqnarray}
${\cal P}_{ij}$ permutes the spins on sites $i$ and $j$.  
The corresponding eigenvalues of (\ref{hamil1}) are obtained from
(\ref{Lambda5}): adding an external magnetic field $H$ and a chemical
potential
\begin{eqnarray}
  && E^{(1)}\left(\{\lambda_j\},\{\nu_\alpha\}\right) -HM^z-\mu N_h
\nonumber\\
  && =  \sum_{k=1}^{N_h+N_\downarrow} \left(H-{2\over\lambda_k^2+1}\right)
     -\sum_{\alpha=1}^{N_h} \left(\mu+{1\over2}H\right) -LH
\label{heig}
\end{eqnarray}
(we have added an external magnetic $H$ field and a (hole) chemical
potential $\mu$ to the Hamiltonian).

To proceed to higher $S$ we iterate the procedure used above: for
$S_1=1/2$ and $S_2=S'-1/2$ arbitrary we use the fact that
$R_{{1\over2}, S'-{1\over2}} (\mu=-iS')\propto{\cal P}_{[S']_+}$ in
the YBE (\ref{ybeSSS}).  This leads to the fusion relation
\begin{equation}
 t^{\{1/2,S\}}\left(\mu-i\left(S'-{1\over2}\right)\right)\,
	    t^{\{S'-1/2,S\}}\left(\mu+{i\over2}\right)
	= t^{\{S',S\}}(\mu) 
	+ t^{\{\left[S'+1/2,S'-1/2\right],S\}}(\mu)\, 
\end{equation}
which allows to determine the eigenvalues of $t^{\{S',S\}}(\mu)$ from
the known ones of $t^{\{1/2,S\}}(\mu)$ and $t^{\{S'-1/2,S\}}(\mu)$ as
\begin{eqnarray}
  \Lambda^{\{S',S\}}\left(\mu|
	\{\lambda_j\}_{j=1}^{N_h+N_\downarrow},
	\{\nu_\alpha\}_{\alpha=1}^{N_h}\right) &=&
  \left( \prod_{k=|S-S'|+1}^{S+S'} {\mu-ik\over\mu+ik} \right)^L
	\prod_{j=1}^{N_h+N_\downarrow}
	{\mu-\lambda_j+iS'\over\mu-\lambda_j-iS'}
\nonumber\\
  && + \Big(\mu-i(S-S')\Big)^L \left\{\ldots\right\}\, .
\end{eqnarray}
The terms in braces are determined by the fusion equations together
with the vanishing of the residues at the simple poles of
$\Lambda^{\{S',S\}} (\mu)$ due to the BAE (\ref{bae3}).  They do not
contribute to the spectrum of the nearest neighbour spin chain
Hamiltonian
\begin{equation}
  {\cal H}^{(S)} = -i \left.{\partial\over\partial\mu} 
	\ln\left(t^{\{S,S\}}(\mu)\right)\right|_{\mu=0} + {\rm const.}
\label{HS}
\end{equation}
whose eigenvalues are
\begin{eqnarray}
  && E^{(S)}\left(\{\lambda_j\},
	\{\nu_\alpha\}\right) -HM^z-\mu N_h
\nonumber\\
  && =  \sum_{k=1}^{N_h+N_\downarrow} \left(H-{2S\over\lambda_k^2+S^2}\right)
     -\sum_{\alpha=1}^{N_h} \left(\mu+{1\over2}H\right) -LH\ .
\label{heigS}
\end{eqnarray}

%
%
\section{Thermodynamic Bethe Ansatz}
\label{sec:tba}
To study the thermodynamics of the doped spin chains (\ref{HS}) we
have to analyze the BAE (\ref{bae3}).  In the thermodynamic limit
$L\to\infty$ general solutions are known to consist of real hole
rapidities $\nu_\alpha$ and complex $n$-\emph{strings} of
spin-rapidities
\begin{equation}
  \lambda_j^{n,k} = \lambda_j^{(n)} + {i\over2}\left(n+1-2k\right)\, ,
  \quad
  k=1,\ldots,n\, .
\label{strings}
\end{equation}
Now we consider solutions of (\ref{bae3}) built from $N_h$ hole
rapidities and $M_n$ $\lambda$-strings of length $n$.  Rewriting the
BAE in terms of the real variables $\nu_\alpha$ and $\lambda_j^n$ and
taking the logarithm we obtain
\begin{equation}
  z_c\left(\nu_\alpha\right) = {I_\alpha\over L}\ ,\qquad
  z_s^{(n)}\left(\lambda_j^{(n)}\right) 
	= {J_j^{(n)}\over L}\ ,
\label{zaehl}
\end{equation}
where $J_j^{(n)}$ and $I_\alpha$ are integers (or half-odd integers)
and the functions $z_i$ are given as
\begin{eqnarray}
  {2\pi}z_s^{(n)}(\lambda) &=& 
	\theta_{n,2S}\left({\lambda}\right)
	-{1\over L}\sum_{m=1}^\infty \sum_{j=1}^{M_{m}} 
	 \Xi_{nm}\left(\lambda-\lambda_j^{(m)}\right)
	+ {1\over L}\sum_{\alpha=1}^{N_h} 
	 \theta_n\left({\lambda-\nu_\alpha}\right)
\nonumber\\
  {2\pi}z_c(\nu) &=&
  {1\over L} \sum_{n=1}^\infty \sum_{j=1}^{M_n}
	\theta_n\left({\nu-\lambda_j^{(n)}}\right)
\label{zz}
\end{eqnarray}
where $\theta_n(x) = 2\arctan(2x/n)$ and 
\begin{eqnarray}
  \theta_{nm}(x) &=& \theta_{m+n-1}\left(x\right)
	+ \theta_{m+n-3}\left(x\right)+\ldots
	+ \theta_{|m-n|+1}\left(x\right)\ ,
\nonumber\\
  \Xi_{nm}(x) &=& \theta_{n+m}\left({x}\right)
 	+2\theta_{n+m-2}\left({x}\right)+\ldots
	+2\theta_{|n-m|+2}\left({x}\right)
	+\left(1-\delta_{nm}\right)\theta_{|n-m|}\left({x}\right)\
 	.
\end{eqnarray}
The quantum numbers $J_j^{(n)}$ and $I_\alpha$ in (\ref{zaehl})
uniquely determine a particular eigenstate of the system.  The
asymptotic behaviour of the functions (\ref{zz}) determine their
possible values.  This allows to introduce densities $\rho(\nu)$ of
the hole rapidities, $\sigma_n(\lambda)$ of the $\lambda$-strings and
the corresponding hole densities $\tilde\rho(\nu)$ and
$\tilde\sigma_n(\lambda)$ with
\begin{equation}
  \sigma_n(x) +\tilde\sigma_n(x) 
	= {\partial\over\partial x} z_s^{(n)}(x)\ ,
\quad
  \rho(x) + \tilde\rho(x) = {\partial\over\partial x} z_c(x)\ .
\end{equation}
In the thermodynamic limit $L\to\infty$ with $N_h/L$ and $M_n/L$ held
fixed these equations become linear integral equations
\begin{eqnarray}
  &&\tilde\sigma_n(x) =
	\left(A_{n,2S}*s\right)(x)
	-\sum_m \left(A_{nm}*\sigma_m\right)(x)
	+\left(a_n*\rho\right)(x)\ ,
\nonumber\\
  &&\rho(x) + \tilde\rho(x) = \sum_n \left(a_n*\sigma_n\right)(x)\ .
\label{intd1}
\end{eqnarray}
Here, $\left(f*g\right)(x)$ denotes a convolution, $2\pi a_n(x) =
\theta_n'(x) = 4n/(4x^2 + n^2)$, $s(x) = 1/(2\cosh\pi
x)$ and
\begin{equation}
  A_{nm}(x) = {1\over2\pi} \Xi_{nm}'(x) + \delta_{nm}\,\delta(x)\ .
\end{equation}
Similarly, the energy (\ref{heigS}) in the thermodynamic limit can be
rewritten in terms of the densities
\begin{equation} 
  E/L = \sum_{n=1}^{\infty} 
	\int{\rm d}x
	\left(\epsilon_n^{(0)}(x)+nH\right)\sigma_n(x)
	 - \int{\rm d}x\left(\mu+{1\over2}H\right)\rho(x)
\end{equation}
where $\epsilon_n^{(0)}(x)=-2\pi\left(A_{n,2S}*s\right)(x)$ are the
bare energies of the $\lambda$-strings.

At finite temperature the equilibrium state is obtained by
minimization of the free energy $F=E-TS$ by variation of $\sigma_n$
and $\rho$.  Here $S$ is the combinatorical entropy \cite{yaya:69}
\begin{eqnarray}
  S/L &=& \sum_{n=1}^\infty \int{\rm d}x\left\{
		\left(\sigma_n+\tilde\sigma_n\right)
		 \ln\left(\sigma_n+\tilde\sigma_n\right)
		-\sigma_n\ln\sigma_n
		-\tilde\sigma_n\ln\tilde\sigma_n\right\}
\nonumber\\
	&&\qquad+ \int{\rm d}x \left\{
		\left(\rho+\tilde\rho\right)
		 \ln\left(\rho+\tilde\rho\right)
		-\rho\ln\rho
		-\tilde\rho\ln\tilde\rho\right\}\ .
\end{eqnarray}
As a result we obtain the thermodynamic Bethe ansatz (TBA) equations
for the energies $\epsilon_n = T\ln(\tilde\sigma_n/\sigma_n)$ of
$\lambda$-strings and $\kappa = T\ln(\tilde\rho/\rho)$ for the hole
rapidities
\begin{eqnarray}
 && \epsilon_n(x) -{T\over2\pi}
	\sum_m \Xi_{nm}'*\ln\left[1+{\rm e}^{-\epsilon_{m}/T}\right](x)
	+ T a_n*\ln\left[1+{\rm e}^{-\kappa/T}\right](x)
  = \epsilon_n^{(0)}(x) +nH
\nonumber\\
 && \kappa(x) 
	+ T \sum_m a_m*\ln\left[1+{\rm e}^{-\epsilon_m/T}\right](x)
  = -\left(\mu+{1\over2}H\right)
\label{tba0}
\end{eqnarray}

An alternative form of these equations can be obtained by using the
identity $\sum_k \left(C_{nk}*A_{km}\right)(x) = \delta_{nm}\,
\delta(x)$ with
\begin{equation}
   C_{nm}(x) =\delta_{nm} \delta(x) -
\left(\delta_{n+1,m}+\delta_{n-1,m}\right)s(x)\ .
\end{equation}
This allows to rewrite the integral eqs.~(\ref{intd1}) as
\begin{eqnarray}
    \delta_{n,2S}\ s(x) &=& \sigma_n(x) +
     \left(C_{nm}*\tilde\sigma_m\right)(x) - 
	\delta_{n,1}\left(s*\rho\right)(x)
\nonumber\\
    \left(a_{2S}*s\right)(x) &=& \tilde\rho(x) + 
	\left([1 + a_2]^{-1}*\rho\right)(x) +
	\left(s*\tilde\sigma_1\right)(x)\ .
\label{intd}
\end{eqnarray}
Similarly, we find for the energy of the corresponding state
\begin{eqnarray} 
  E/L &=& E_0^{(S)}/L 
	-\int{\rm d}x \left[2\pi(a_{2S}*s)+\mu\right]\rho(x)
      +\int{\rm d} x 2\pi s(x) \tilde\sigma_{2S}(x)
\\
 &&	- \lim_{n\rightarrow\infty}Hn\int {\rm d} x
	   \tilde\sigma_n(x)\ 
\nonumber
\end{eqnarray}
where $E_0^{(S)}$ is the ground state energy of the spin-$S$
Takhtajan--Babujian chain in a vanishing magnetic field \cite{babu:83}
\begin{equation}
  E_0^{(S)} = \left\{
   \begin{array}{ll}
	-\sum_{k=1}^S {2\over 2k-1} & \hbox{for~integer}\,S\\[4pt]
	-2\ln2-\sum_{k=1}^{S-1/2} {1\over k} & \hbox{for~half-integer}\,S
   \end{array}
\right.
\end{equation}
Finally, an equivalent form of the TBA equations (\ref{tba0}) is
\begin{eqnarray}
  \epsilon_n(x) &=&
	T\left(s*\ln\left[1 + {\rm e}^{\epsilon_{n -1}/T}\right]
	      \left[1 + {\rm e}^{\epsilon_{n +1}/T}\right]\right)(x) 
\nonumber\\
&&\qquad
	- 2\pi \delta_{n,2S}\,s(x) 
	- \delta_{n,1}T
	  \left(s*\ln\left[1 + {\rm e}^{- \kappa/T}\right]\right)(x)\ ,
\label{tba1}
\end{eqnarray}
subject to the condition $\lim_{n\to\infty}(\epsilon_n/n) = H$ and
\begin{equation}
  -[ 2\pi a_{2S}*s(x) + \mu] 
	- T\left(s*\ln\left[1 + {\rm e}^{\epsilon_{1}/T}\right]\right)(x) 
   = \kappa(x) 
	+ T\left(R*\ln\left[1 + {\rm e}^{- \kappa/T}\right]\right)(x)\
\label{kappa}
\end{equation}
where $R = a_2*(1 + a_2)^{-1}$.

In terms of the solutions to these equations the free energy reads
\begin{equation}
  F/L = E_0^{(S)}/L
	-T\int{\rm d}x s(x)
	 \ln\left[1 + {\rm e}^{\epsilon_{2S}(x)/T}\right] 
	-T\int{\rm d}x (a_{2S}*s)(x)\,
	 \ln\left[1 + {\rm e}^{-\kappa(x)/T}\right]\ .
\label{FreeE}
\end{equation}

%
%

\section{Zero temperature phases in a magnetic field}
\label{sec:phasH}
In the limit $T\to0$ the TBA eqs.\ (\ref{tba0}) become linear integral
equations.  As a consequence of (\ref{tba1}) only $\epsilon_1(x)$,
$\epsilon_{2S}(x)$ and $\kappa(x)$ can take negative values for
certain $x$.  Hence we have to solve three coupled integral equations
for these quantities which in turn determine  all other dressed energies
\begin{eqnarray}
  &&  \epsilon_n(x) + {1\over2\pi}\left\{
	\Xi_{n1}'*\epsilon_1^{(-)} 
	+ \Xi_{n,2S}'*\epsilon_{2S}^{(-)}\right\}(x)
	- a_n*\kappa^{(-)}(x)
  = \epsilon_n^{(0)}(x) +nH
\nonumber\\
  &&  \kappa(x) - \left\{ a_1*\epsilon_1^{(-)} 
			+ a_{2S}*\epsilon_{2S}^{(-)}\right\}(x)
  = -\left(\mu+{1\over2}H\right)\ ,
\label{dressE}
\end{eqnarray}
where $f^{(\pm)}(x) = \theta(\pm f(x)) f(x)$.

To discuss the solutions of these equations further we have to
distinguish various cases:
\subsection{$\mu>H/2$}
In this regime we have $\kappa(x)<0$ and $\epsilon_1(x)<0$ for all
$x$.  This allows to express these functions in terms of the remaining
unknown function $\epsilon_{2S}^{-}(x)$.  From (\ref{dressE}) we find
\begin{eqnarray}
 && \kappa(x) = -2\mu+\left(a_1*\epsilon_1^{(0)}\right)(x)
\nonumber\\
 && \epsilon_1(x) = \epsilon_1^{(0)}(x) - \left(\mu-{H\over2}\right)
	-\left(a_{2S-1}*\epsilon_{2S}^{(-)}\right)(x)
\nonumber\\
 && \epsilon_{n>1}(x) + {1\over2\pi}\left(
	\Xi_{n-1,2S-1}'*\epsilon_{2S}^{(-)}\right)(x)
	= -2\pi\left(A_{n-1,2S-1}*s\right)(x)+(n-1)H
\nonumber
\end{eqnarray}
The last set of equations can be identified with the integral
equations for the dressed energies of the spin-$S-1/2$
Takhtajan--Babujian model, hence this regime corresponds to the
completely doped case (i.e.\ $N_h=L$ holes).  For magnetic field
$H>H^{(S-1/2)}$ with
\begin{equation}
  H^{(\sigma)}>{2\over\sigma}\sum_{k=1}^{2\sigma}{1\over2k-1}
\label{Hsat1}
\end{equation}
the system is in a ferromagnetically saturated state with maximal
magnetization $M^z=L(S-1/2)$.

\subsection{$-H/2<\mu<H/2$}
Here we find from (\ref{dressE}) that $\kappa(x)\equiv
\kappa^{(-)}(x)<0$ for all $x$ while $\epsilon_1(x)$ can take
non-negative values.  Eliminating $\kappa(x)$ from the integral
equations for $\epsilon_1$ and $\epsilon_{2S}$ we obtain
\begin{eqnarray}
  && \epsilon_1(x) + \left\{a_{2S-1}*\epsilon_{2S}^{(-)}\right\}(x)
	= \epsilon_1^{(0)}(x) - \mu+{1\over2}H\,,
\nonumber\\
  && \epsilon_{2S}(x) + \left\{a_{2S-1}*\epsilon_{1}^{(-)}
		      + 2\sum_{k=1}^{2S-1}a_{2k}*\epsilon_{2S}^{(-)}
			\right\}(x)
 = \epsilon_{2S}^{(0)}(x) - \mu+{4S-1\over2}H\, .
\nonumber
\end{eqnarray}
In this regime we find $\epsilon_{1}>0$ and $\epsilon_{2S}>0$ for
\begin{equation}
  \mu< \min\left\{{4S-1\over2}H -4\sum_{k=1}^{2S}{1\over2k-1}\, ,\,
		  {1\over2}H-{2\over S}\right\}\ .
\label{ferro1}
\end{equation}
Positive dressed energies for the $\lambda$-strings imply $M_n=0$ and
from (\ref{intd1}) we find that $N_h=0$ in this region.  Hence for
$T\to 0$ (\ref{ferro1}) belongs to the ferromagnetically saturated
phase of the undoped system, namely the spin-$S$ Takhtajan-Babujian
model.  This phase exists for magnetic fields $H>H^{(S)}$.

Increasing the hole chemical potential to values $\mu>H/2-2/S$ holes
are added but the ground state continues to be fully polarized: For
the dressed energies this corresponds to $\epsilon_{2S}>0$ and while
the real spin rapidities fill all states with negative
$\epsilon_1(x)$.  As a consequence of the free fermionic nature of
this state these dressed energies can be expressed in terms of their
free values
\begin{equation}
  \epsilon_{1}(x) = \epsilon_1^{(0)}(x) -\left(\mu-{H\over2}\right)\ .
\end{equation}
The lower boundary of this phase in the $\mu$--$H$ plane is given by
the condition $\min_x\left\{\epsilon_{2S}(x)\right\}=0$.

For magnetic fields $(4S-1/2)H<\mu+4\sum_{k=1}^{2S}1/(2k-1)$ the
ground state is a filled sea of  $\lambda$-strings of
length $2S$ with negative energy
\begin{equation}
  \epsilon_{2S}(x)
	+ \left\{\sum_{k=1}^{2S-1} 
		a_{2k}*\epsilon_{2S}^{(-)}\right\}(x)
	= \epsilon_{2S}^{(0)}(x) + \left(2S-{1\over2}\right)H -\mu\ .
\label{phaseX}
\end{equation}
The other dressed energies $\kappa(x)<0$ and $\epsilon_{n\ne2S}(x)>0$
can be expressed in terms of the solution of this equation.  In this
region of parameters the system has a finite concentration of holes
\emph{and} overturned spins.  Although one might na\"{\i}vely expect
two types of massless excitations in such situation only one such
branch with dispersion (\ref{phaseX}) is realized in this system which
turns out to describe the charge excitations.  Hence spin excitations
are gapped in this regime \cite{frso:pp}.
\subsection{$\mu<-H/2$}
\label{ssec:phas3}
Again we find several phases that can be characterized by the
configurations of strings present in the ground state: For $H>H^{(S)}$
all dressed energies are positive corresponding to completely
polarized undoped state.

For smaller magnetic fields $\epsilon_{2S}(x)$ takes negative values
in some interval to be determined from
\begin{equation}
 \epsilon_n(x) + {1\over2\pi}\left\{
	\Xi_{n,2S}'*\epsilon_{2S}^{(-)}\right\}(x)
  = \epsilon_n^{(0)}(x) +nH\ .
\end{equation}
These are the Bethe ansatz equations of the spin-$S$
Takhtajan--Babujian chain.  This phase becomes unstable against hole
creation for chemical potentials
\begin{equation}
  \mu>\int {\rm d}a_{2S}(x)\epsilon_{2S}^{(-)}(x) -{1\over2}H
	\to \psi\left({2S+1\over4}\right)
	   -\psi\left({2S+3\over4}\right) \mathrm{~for~}H=0
\end{equation}
($\psi(x)$ is the digamma function).  Beyond this line the ground
state is built from a filled sea of $\lambda$-strings with energies
$\epsilon_{2S}<0$ and another sea of charge rapidities with energies
$\kappa(x)<0$.
Increasing the chemical potential further negative energy solutions
for the real spin rapidities $\epsilon_1$ appear giving rise to a
third condensate of Bethe rapidities.

\begin{figure}[ht]
\begin{center}
\leavevmode
\epsfxsize=0.65\textwidth
(a)\epsfbox{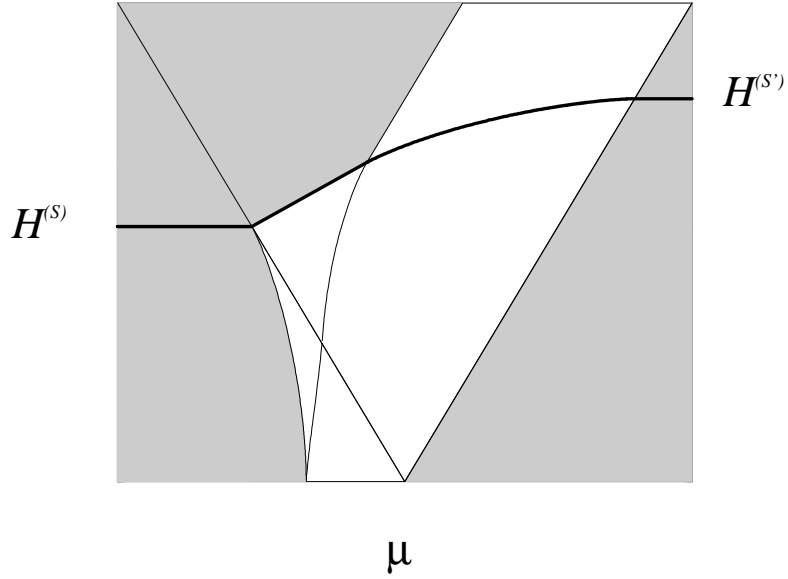}
\end{center}
\begin{center}
\leavevmode
\epsfxsize=0.65\textwidth
(b)\epsfbox{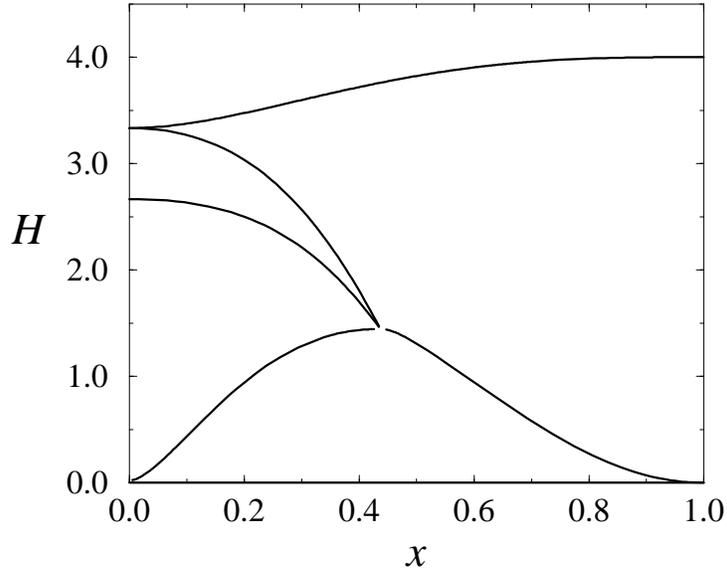}
\end{center}
\caption{(a) Schematic phase diagram of the doped spin chain in the
$\mu$--$H$ plane: The bold line denotes the transition to a fully
polarized state, interpolating between the saturation fields
(\protect{\ref{Hsat1}}) for the spin $S$ and $S'=S-1/2$
Takhtajan--Babujian chains.  The left (right) shaded region
corresponds to the undoped (completely doped) regime.  (b) Phase
diagram of the doped $S=1$ chain as a function of hole concentration
$x$.
\label{fig:phasemu}}
\end{figure}

From the cases considered above we obtain the qualitative zero
temperature phase diagram of the doped spin-$S$ system in the
$\mu$--$H$ plane presented in Fig.~\ref{fig:phasemu}(a).  Using
Eqs.~(\ref{intd1}) the corresponding phase boundaries can be given as
a function of the hole concentration $x=N_h/L$.  For $S=1$ this is
shown in Fig.~\ref{fig:phasemu}(b).

\section{Low-temperature thermodynamics at $H=0$}
\label{sec:phasT}
Further simplification is possible in the case of finite doping in a
vanishing magnetic field which corresponds to chemical potentials $\mu
\in \left[\psi((2S+1)/4)-\psi((2S+3)/4),0\right]$.  Furthermore,
$\epsilon_1(x)<0$, $\epsilon_{2S}(x)<0$ for all $x$ while
$\kappa(x)<0$ for $|x| < Q$ in this regime.  All other dressed
energies vanish in this limit.  Eliminating the $\epsilon_n$ from the
equation for the energy of the holes we obtain
\begin{equation}
  -[ 2\pi a_{2S}*s(x) + \mu] = \kappa(x) -\int_{-Q}^{Q}{\rm d}y
   R(x-y)\kappa(y)\
\label{kappa0}
\end{equation}
where the Fermi point $Q$ is determined by the condition $\kappa(\pm
Q)=0$.  Similarly, the density of hole rapidities $\rho_0(x)$ in this
regime is given by
\begin{eqnarray}
  \rho_0(x) - \int_{-Q}^Q {\rm d} yR(x - y)\rho_0(y) = a_{2S}*s(x)\ .
\end{eqnarray}
Excitations with charge rapidities near $\pm Q$ are massless.  The
velocity of this charge mode can be obtained from the dispersion
(\ref{kappa0}) 
\begin{equation}
  v = \frac{1}{2\pi\rho_0(Q)}
	\left.\frac{\partial\kappa_0}{\partial x}\right|_{x = Q}\ .
\label{v}
\end{equation}
Similarly, one has massless excitations near $x=\pm\infty$ in the
magnetic sector with energies $\epsilon_{1}(x)$ and $\epsilon_{2S}(x)$
with velocities
\begin{equation}
  v_{2S} = \lim_{x\to\infty}
	{\epsilon_{2S}'(x)\over2\pi\sigma_{2S}(x)} \equiv \pi\ ,
\quad
  v_{1} = \lim_{x\to\infty}
	{\epsilon_{1}'(x)\over2\pi\sigma_1(x)}
	= -{1\over2}\,
        \frac{\int_{-Q}^Q{\rm d} y {\rm e}^{\pi y}\kappa_0(y)}{
                \int_{-Q}^Q{\rm d} y{\rm e}^{\pi y}\rho_0(y)}\ .
\label{veloS}
\end{equation}

As a consequence of the behaviour of the dressed energies as $H\to0$
we can replace $\kappa$ in Eq.~(\ref{tba1}) by its zero temperature
value $\kappa_0(x)$ and the driving terms by their asymptotics to
obtain the leading low temperature behaviour.  As a result we get
\begin{equation}
  \epsilon_n(x) = Ts*\ln[1 + {\rm e}^{\epsilon_{n -1}(x)/T}][1 +
  {\rm e}^{\epsilon_{n +1}(x)/T}]
  - 2\pi\delta_{n,2S}\,{\rm e}^{- \pi|x|} 
  - 2\pi A\delta_{n,1}{\rm e}^{- \pi|x|}
\label{tba2}
\end{equation}
where $2\pi A = - \int_{-Q}^Q{\rm d} y {\rm e}^{\pi y}\kappa_0(y)$.

To move further we have to separate the contributions to the free
energy stemming from the charge-sector from those due to the
$\epsilon_n$.  Considering low temperatures again the leading
contributions to $\kappa$ come from the vicinity of the Fermi wave
vectors $\pm Q$. In this region one can safely neglect contributions
to Eq.~(\ref{kappa}) from $\epsilon_1$ and rewrite it as
\begin{eqnarray}
 && -[2\pi a_{2S}*s(x) + \mu] - TR*\ln[1 + {\rm e}^{- |\kappa(x)|/T}]
\nonumber\\
 &&\qquad = \kappa(x) -  \int_{-Q}^Q {\rm d} yR(x - y)\kappa(y)
\end{eqnarray}
where $Q$ is determined by the condition $\kappa(\pm Q) = 0$. 
Using the procedure introduced by Takahashi \cite{taka:71b}, we can
rewrite the free energy (\ref{FreeE}) as $F/L = E_0^{(S)}/L + f_c +
f_s$ where
\begin{eqnarray}
  f_c &=& - T\int{\rm d}x \rho_0(x)
	\ln\left[1 + {\rm e}^{- |\kappa_0(x)|/T}\right]
	  \approx- \pi T^2/6v\ ,
\label{Free}\\
  f_s &=& - T\int{\rm d}x s(x)
	\ln\left[1 + {\rm e}^{\epsilon_{2S}(x)/T}\right] 
          - T\int{\rm d}x (s*\rho_0)(x)
	\ln\left[1 + {\rm e}^{\epsilon_1(x)/T}\right]\ .
\label{free2}
\end{eqnarray}
Now the thermodynamics is described by Eq.~(\ref{Free}) representing a
scalar bosonic mode (the charge sector) and by Eqs.~(\ref{tba2}) and
(\ref{free2}) for the spin sector. 

At low temperatures the spin contribution $f_s$ is dominated by
contributions from the regions $v_{2S}\exp(- \pi |x|) \sim T$ where
$|\epsilon_{2S}| \sim T$ and the second one by the regions $v_1\exp(-
\pi |x|) \sim T$.  The leading temperature dependence of $f_s$ at low
$T$ can be obtained by rewriting (\ref{tba2}) for large positive $x$
as
\begin{equation}
  \varphi_n(x) = s*\ln[1 + {\rm e}^{\varphi_{n -1}}][1 +
  {\rm e}^{\varphi_{n+1}}]
  - \delta_{n,2S}\,{\rm e}^{- \pi x} 
  - A\delta_{n,1}{\rm e}^{- \pi x}
\label{tba22}
\end{equation}
in terms of the $T$-independent functions
\[
  \varphi_n(x) = {1\over T} 
	\epsilon_n\left(x-{1\over\pi}\ln{T\over2\pi}\right)\ .
\]
In the low-$T$ limit we can replace $s(x)$ and $s*\rho_0(x)$ in
(\ref{free2}) by their asymptotics to obtain the free energy
\begin{equation}
  f_s \simeq - {\pi T^2\over6} \left(  {c_{2S}\over v_{2S}}
				+ { c_1 \over v_1} \right)\ .
\label{free22}
\end{equation}
Such an expression is characteristic of a system two with massless
excitations with velocities (\ref{veloS}).  In cases where these
excitations can be characterized by different \emph{observable}
quantum numbers the coefficients $c_i$ are the central charges of the
underlying Virasoro-algebra thus determining the universality class of
the system.  In this case they are given in terms of the solutions of
(\ref{tba22}) by
\begin{equation}
  c_{2S} = {6\over\pi} \int{\rm d}x\, {\rm e}^{-\pi x}
		\ln\left[1 + {\rm e}^{\varphi_{2S}(x)}\right]\ ,
\qquad
  c_1 = {6\over\pi}\int{\rm d}x\, (A {\rm e}^{-\pi x})
	\ln\left[1 + {\rm e}^{\varphi_1(x)}\right]\ .
\label{charges}
\end{equation}
Using standard methods \cite{babu:83,klme:90,zamo:91} for the analysis
of the TBA equations we find that their sum can be written as
\begin{equation}
  c_{2S}+c_{1} = {6\over\pi^2} \sum_n \left[
    {\cal L}\left({{\rm e}^{\varphi_n(x)}
		\over1+{\rm e}^{\varphi_n(x)}}\right)
  \right]_{x=-\infty}^\infty
\label{csum}
\end{equation}
where ${\cal L}(x)$ is Rogers dilogarithm function
\[
   {\cal L}(x) = -{1\over2}\int_0^x {\rm d}t\,
	\left[ {\ln t\over 1-t} + {\ln(1-t)\over t}\right]\ .
\]
Hence, the $c_{2S}+c_1$ is completely determined by the asymptotic
behaviour of the solutions of (\ref{tba22}) as $x\to\pm\infty$:
\begin{eqnarray}
  \lim_{x\to\infty}\varphi_n(x) &=& \ln\left((n+1)^2-1\right)\ ,
\nonumber\\
  \lim_{x\to-\infty}\varphi_n(x) &=&
	\left\{\begin{array}{ll}
	 \ln\left((n-2S+1)^2-1\right) &	{\rm for~}n>2S \\
  	 \ln\left({\sin^2(\pi n/2S+1)\over\sin^2(\pi/2S+1)} -1\right) &
		{\rm for~}1<n<2S\\
	 -\infty & {\rm for~} n=1,\,2S
	       \end{array}
\right.\ ,
\nonumber
\end{eqnarray}
giving
\begin{equation}
   c_{2S}+c_1 = 2 {4S-1\over2S+1}\ 
\label{SumC}
\end{equation}
independent of the doping (i.e.\ $A$).  

The individual values of the $c_i$ are easily calculated for small and
large doping corresponding to $A\to0$ and $A\to\infty$, respectively.
In these cases the regions contributing to the integrals
(\ref{charges}) are well separated and the functions $\varphi_n(x)$
take constant values in between.  For small doping ($A\ll1$) we find
$\varphi_n(x)=\varphi_n^{(0)}$ for $\ln A\ll \pi x\ll 0$ with
\begin{equation}
	\varphi_n^{(0)} =
	\left\{\begin{array}{ll}
	 \ln\left((n-2S+1)^2-1\right) &	{\rm for~}n>2S \\
	 -\infty & {\rm for~} n=2S\\
  	 \ln\left({\sin^2(\pi n/2(S+1))\over\sin^2(\pi/2(S+1))} -1\right) &
		{\rm for~}1\le n<2S
	       \end{array}
   \right.\, .
\end{equation}
Hence, near $x\approx0$ they behave as in the undoped system giving
the central charge $3S/(S+1)$ of the $SU(2)_{2S}$ WZNW model.  In the
region around $x\approx \ln A$ the $\varphi_{n<2S}$ are solutions of
the {\em finite} set of TBA of the minimal unitary model ${\cal
M}_{p}$ \cite{zamo:91} with central charge $c_1=1-6/(p(p+1))$ where
$p=2S+1$ (this is the Ising model for $S=1$ (see \cite{fpt:98}),
tricitical Ising model for $S=3/2$, three-state Potts model for $S=2$,
tricritical three state Potts model for $S=5/2$ and so forth).
Putting everything together we find the leading contribution to the
spin part (\ref{free2}) to the free energy at small doping
\begin{equation}
  f_s = -{\pi T^2\over 6v_{2S}}\, {3S\over S+1}
	-{\pi T^2\over 6v_1}\left\{1-{3\over{(S+1)(2S+1)}}\right\}\, .
\label{fs_0}
\end{equation}

Proceeding in an analogeous way in the limit of large doping ($A\gg1$)
corresponding to a spin-$(S-1/2)$ chain doped with spin-$S$ carriers
we find
\begin{equation}
	\varphi_n^{(0)} =
	\left\{\begin{array}{ll}
	 \ln\left(n^2-1\right) &	{\rm for~}n>1 \\
	 -\infty & {\rm for~} n=1\\
	       \end{array}
   \right.\,
\end{equation}
for $0\ll \pi x\ll \ln A$.  In this limit the low temperature
contributions to $f_s$ can be written as the sum of a $SU(2)_1$ and a
$SU(2)_{2S-1}$ WZNW model, the latter being the well known continuum
limit of the pure spin-$(S-1/2)$ Takhtajan-Babujian model:
\begin{equation}
  f_s = -{\pi T^2\over 6v_{2S}}\, {6S-3\over 2S+1}
	-{\pi T^2\over 6v_1}\ .
\label{fs_1}
\end{equation}

\begin{figure}[ht]
\begin{center}
\leavevmode
\epsfxsize=0.65\textwidth
(a)\epsfbox{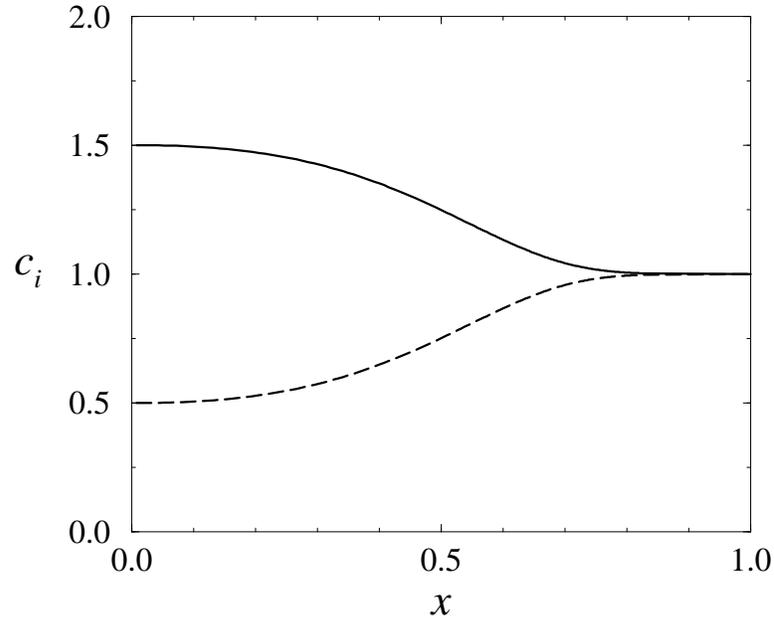}
\end{center}
\begin{center}
\leavevmode
\epsfxsize=0.65\textwidth
(b)\epsfbox{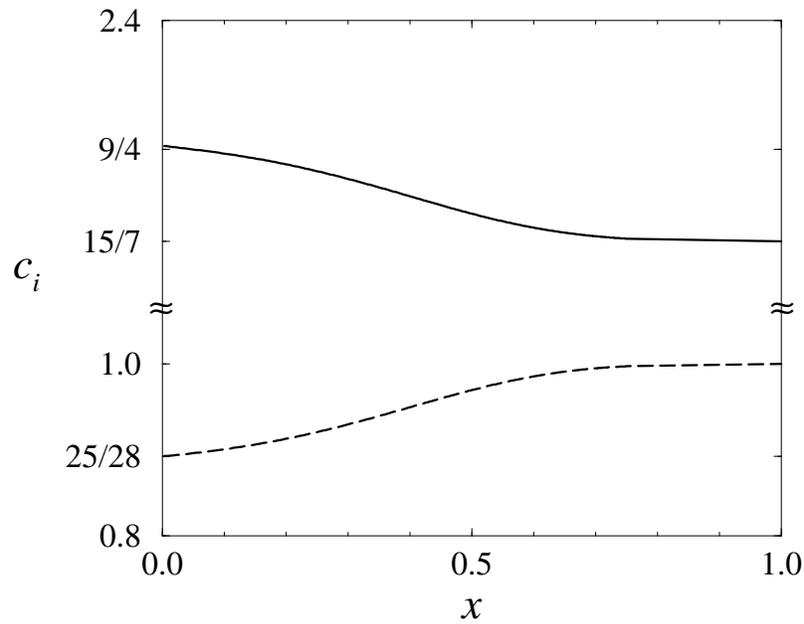}
\end{center}
\caption{Dependency of $c_1$ (dashed line) and $c_{2S}$ (full line)
on the concentration $x$ of holes for (a) $S=1$, (b) $S=3$.
\label{fig:ccx}}
\end{figure}

For finite values of $A$ the coefficients $c_{2S}$ and $c_1$ in
(\ref{free22}) have to be determined numerically.  They are found to
interpolate smoothly between their limiting values in (\ref{fs_0}) and
(\ref{fs_1}).  For $S=1$ and $S=3$ their doping dependence is shown in
Figure~\ref{fig:ccx}.

\section{Summary and Conclusion}
To summarize, we have introduced a class of integrable models
describing a magnetic system which upon doping interpolates between
the integrable spin-$S$ and $S-1/2$ Takhtajan-Babujian chains.  These
models arise when considering vertex models invariant under the action
of the graded Lie algebra $gl(2|1)$ with the local quantum spaces
carrying the `atypical' higher-spin representations
$\left[S\right]_+$.  Their solution by means of the algrebraic Bethe
Ansatz allows for a detailed study of their low temperature phase
diagram.  The spectrum of low lying excitations is described in terms
of the dressed energies satisfying the TBA equations (\ref{tba1}) and
(\ref{kappa}).  Without an external magnetic field the critical
degrees of freedom separate into charge and magnetic modes as is well
known in the Tomonaga-Luttinger liquid models for one-dimensional
correlated electrons (see e.g.\ Refs.~\cite{frko:90,kaya:91}).
Different from these models, however, one finds \emph{two} branches of
low lying modes in the magnetic sector which at small (large) doping
can be identified with higher level $SU(2)_k$ WZW models and a minimal
model (free boson).  The WZW models have to be present in order to
reproduce the well understood critical behaviour of the undoped and
completely doped limiting cases.  The second gapless magnetic mode,
however, is quite peculiar: its appearence in the low energy of the
undoped system is crucial to allow for the smooth crossover between
the limiting cases (\ref{fs_0}) and (\ref{fs_1}) subject to the
constraint $c_{2S}+c_1={\rm const}$.

The low-$T$ behaviour of the $S=1$ integrable model has motivated the
proposition of an effective field theory of four (real) Majorana
fermions as a possible starting point for studies of perturbations
around the integrable model \cite{fpt:98}.  While free field
representations could be used for the constituents of the
\emph{undoped} model, interaction terms between the two sectors had to
be introduced to reproduce the change of the coefficients $c_{2S}$ and
$c_1$ with the hole concentration observed in the exact solution.  The
possible form of this interaction term is constrained by the
$SU(2)$-symmetry of the model without a magnetic field.
%
%
A similar construction of an $SU(2)$-invariant effective low energy
field theory for the $S>1$ models introduced here is possible by using
the fact that the minimal models can be obtained within a GKO coset
construction applied to \cite{gko:85,gko:86}
\begin{equation}
  {SU(2)_{2S-1}\otimes SU(2)_{1} \over SU(2)_{2S}}\ .
\label{GKO}
\end{equation}
In fact, the observed change in the conformal weights attributed to
the magnetic modes between the limiting cases of the undoped and the
completely doped system appear to be just a `adiabatic' realization of
this construction
\begin{equation}
  SU(2)_{2S} \otimes {\cal M}_{2S+1} \longrightarrow
    SU(2)_{2S-1}\otimes SU(2)_{1}\ .
\end{equation}
On the other hand, taking the limit $H\to 0$ starting from the phase
discussed in Section~\ref{ssec:phas3} one may obtain a different field
theoretical description of the $SU(2)$-symmetric phase: There the
critical degrees of freedom can be described in terms of two free
bosons each contributing $c=1$ to the sum $c_{2S}+c_1$.  For $S=1$
this should give a complete description of the massless magnetic
modes.  It is likely that the apparent difference between the $H\to0$
limit and the $H=0$ model can be understood as a rotation in the space
of the effective fields (note that no \emph{physical} field couples to
one of the magnetic modes alone) \cite{fab:pc}.  For $S>1$ one has
$c_{2S}+c_1>2$ from (\ref{SumC}).  Here, the difference to the finite
field critical properties is similar to the one observed in the
Takhtajan-Babujian models \cite{tsve:88}: it is due to the appearence
of gap for parafermionic degrees of freedom in the critical theory for
any non-zero magnetic field.

\section*{Acknowledgements}
I am grateful to A.~M.\ Tsvelik and F.~H.~L.\ E\ss{}ler for important
discussions.  This work has been supported by the Deutsche
Forschungsgemeinschaft under Grant No.\ Fr~737/2--3.

\newpage
\appendix
\section{Two more Bethe Ans\"atze}
\label{app:ba}
As mentioned above, the grading of the underlying algebra leads to
different (though equivalent) sets of Bethe Ansatz equations (BAE)
when using different orderings of the basis  \cite{kulish:85}.

A second Bethe Ansatz can be obtained by starting from the pseudo
vacuum $|\Omega_S\rangle$ by choosing a different reference state in
the auxiliary eigenvalue problem for the amplitudes $F^{a_n\cdots
a_1}$.  Following Refs.~\cite{kulish:85,esko:92,foka:93} the
roots of
\begin{eqnarray}
   \left( {\lambda_j+iS\over \lambda_j-iS} \right)^L =
   \prod_{\alpha=1}^{N_\downarrow}
	{{\lambda_j-\nu_\alpha+{i\over2}} \over
	 {\lambda_j-\nu_\alpha-{i\over2}} },\ &&
	j=1,\ldots,N_h+N_\downarrow\ ,
\label{bae2}\\
   1 = \prod_{k=1}^{N_h+N_\downarrow}
	{{\nu_\alpha - \lambda_k +{i\over2}} \over
	 {\nu_\alpha - \lambda_k -{i\over2}}},\ &&
    \alpha=1,\ldots,N_\downarrow\ .
\nonumber
\end{eqnarray}
are found to parametrize eigenstates of the transfer matrix
(\ref{trans3S}) with eigenvalues ($\tilde\lambda_j\equiv
\lambda_j-iS$, $\tilde\nu_\alpha\equiv \nu_\alpha-iS-i/2$)
\begin{eqnarray}
  &&\Lambda_{3}\left(\mu|
	\{\tilde\lambda_j\}_{j=1}^{N_h+N_\downarrow},
	\{\tilde\nu_\alpha\}_{\alpha=1}^{N_h}\right) =
  {\mu}^L \prod_{j=1}^{N_h+N_\downarrow}
	{\mu-\tilde\lambda_j+i \over \mu-\tilde\lambda_j}
\nonumber\\
 &&\qquad
 +\left({\mu+2iS}\right)^L\prod_{\alpha=1}^{N_\downarrow}
	{\tilde\nu_\alpha-\mu +i \over \tilde\nu_\alpha-\mu}
  \left\{ 1 - \prod_{j=1}^{N_h+N_\downarrow}
	{\mu-\tilde\lambda_j+i \over \mu-\tilde\lambda_j} \right\}\ .
\end{eqnarray}
(The spectrum of the other transfer matrices defined in
(\ref{transSSS}) can be obtained from this expression by means of the
fusion equations derived in Section~\ref{sec:fusion}).

Alternatively, we may start from the fully polarized state
$|\Omega_{S-{1\over2}}\rangle = \prod_{n=1}^L |S+{1\over2},
S-{1\over2}, S-{1\over2}\rangle_n$ with maximal number of
spin-$\left(S-{1\over2}\right)$-holes in the system.  Now, eigenstates
of (\ref{trans3S}) obtained by adding $N_e$ particles to
$|\Omega_{S-{1\over2}}\rangle$ and lowering the spin by $N_\downarrow$
are parametrized by solutions of the third set of BAE
\cite{pfannm:diss}
\begin{eqnarray}
   \left( {\lambda_j+iS\over \lambda_j-iS} \right)^L &=&
	\prod_{\alpha=1}^{N_\downarrow}
	{{\lambda_j-\nu_\alpha +{i\over2}} \over
	 {\lambda_j-\nu_\alpha -{i\over2}} } ,
   \quad j=1,\ldots,N_e\ ,
\label{bae1}\\
   \left(\nu_\alpha+i(S-{1\over2}) \over 
	 \nu_\alpha-i(S-{1\over2}) \right)^L &=&
   \prod_{\beta\ne\alpha}^{N_{\downarrow}}
	{{\nu_\alpha-\nu_\beta+i} \over {\nu_\alpha-\nu_\beta-i}}\,
   \prod_{j=1}^{N_e} {{\nu_\alpha-\lambda_j-{i\over2}} \over
		      {\nu_\alpha-\lambda_j+{i\over2}}} ,
\nonumber\\
    &&   \quad \alpha=1,\ldots,N_\downarrow\ .
\nonumber
\end{eqnarray}
The corresponding eigenvalues are ($\tilde\lambda_j\equiv
\lambda_j-iS-i$, $\tilde\nu_\alpha\equiv \nu_\alpha-iS-i/2$)
\begin{eqnarray}
  &&\Lambda_{3}\left(\mu|
	\{\tilde\lambda_j\}_{j=1}^{N_e},
	\{\tilde\nu_\alpha\}_{\alpha=1}^{N_h}\right) =
 -\left({\mu+2iS+i}\right)^L \prod_{j=1}^{N_e}
	{\tilde\lambda_j-\mu+i \over \tilde\lambda_j-\mu}
\nonumber\\
 &&\qquad
 +\left({\mu+2iS}\right)^L\prod_{\alpha=1}^{N_\downarrow}
	{\tilde\nu_\alpha-\mu +i \over \tilde\nu_\alpha-\mu}
 +\left({\mu+i}\right)^L\prod_{j=1}^{N_e}
	{\tilde\lambda_j-\mu+i \over \tilde\lambda_j-\mu}
  \prod_{\alpha=1}^{N_\downarrow}
	{\mu-\tilde\nu_\alpha +i \over \mu-\tilde\nu_\alpha}\ .
\end{eqnarray}
For $S={1\over2}$ Eqs.~(\ref{bae1}) become Lai's BAE for the
supersymmetric $t$--$J$ model \cite{lai:74,schl:87}.

\section{Equivalence of the Bethe Ans\"atze}
\label{app:eq}
In this appendix the equivalence of the sets (\ref{bae3}) and
(\ref{bae1}) of Bethe Ansatz equations starting from the
fully polarized state of spin-$S$ and $S-{1/2}$ multiplets
respectively is shown by means of a particle-hole transformation in
the space of the rapidities.

Following Refs.~\cite{woyn:83,bcfh:92} we rewrite the second set of
Eqs.~(\ref{bae3}) as $P(\nu_\alpha)=0$ with the polynomial
\begin{equation}
  P(\omega) = 
  \prod_{k=1}^{N_h+N_\downarrow}
  \left(\omega-\lambda_k-\frac{i}{2}\right)
 -\prod_{k=1}^{N_h+N_\downarrow}
  \left(\omega-\lambda_k+\frac{i}{2}\right)\ .
\label{app:Pdef}
\end{equation}
According to (\ref{bae3}) the first $N_h$ of the $N_h+N_\downarrow$
roots of $P(\omega)$ can be identified as $\omega_\alpha=\nu_\alpha$,
$\alpha=1,\ldots N_h$.  Labelling the remaining $N_\downarrow$ ones as
$\nu_{\alpha}'$ we have
\begin{eqnarray}
  \sum_{\alpha=1}^{N_h} 
	\ln\left({\lambda_\ell-\nu_\alpha+{i\over2}\over
		  \lambda_\ell-\nu_\alpha-{i\over2}}\right)
 &=& \sum_{\alpha=1}^{N_h} {1\over2\pi i}\oint_{C_\alpha}{\rm d}z
	\ln\left({\lambda_\ell-z+{i\over2}\over
		  \lambda_\ell-z-{i\over2}}\right)
	{{\rm d}\over{\rm d}z} \ln P(z)
\nonumber\\
 &=& -\sum_{\alpha=1}^{N_\downarrow}
	\ln\left({\lambda_\ell-\nu_\alpha'+{i\over2}\over
		  \lambda_\ell-\nu_\alpha'-{i\over2}}\right)
     + \ln\left(P(z_n)/P(z_p)\right)\ ,
\label{app:contour}
\end{eqnarray}
where $C_\alpha$ is a contour enclosing $\nu_\alpha$ and
$z_{n,p}=\lambda_\ell\pm{i/2}$ are the end points of the branch cut of
the logarithm in (\ref{app:contour}).  From the definition
(\ref{app:Pdef}) we have
\begin{equation}
  P(\lambda_\ell\pm{i\over2}) = \mp \prod_{k=1}^{N_h+N_\downarrow}
	\left(\lambda_\ell-\lambda_k\pm i\right)\ .
\end{equation}
which -- when used in (\ref{app:contour}) -- implies that
\begin{equation}
   \prod_{\alpha=1}^{N_h} 
	{\lambda_\ell-\nu_\alpha+{i\over2}\over
	 \lambda_\ell-\nu_\alpha-{i\over2}} =
  -\prod_{\alpha=1}^{N_\downarrow}
	{\lambda_\ell-\nu_\alpha'-{i\over2}\over
	 \lambda_\ell-\nu_\alpha'+{i\over2}}\,
   \prod_{k=1}^{N_h+N_\downarrow}
	{\lambda_\ell-\lambda_k+i \over \lambda_\ell-\lambda_k-i}\ .
\end{equation}
Using this identity in the first of Eqs.~(\ref{bae3}) we obtain
\begin{equation}
   \left(\lambda_\ell +iS\over  \lambda_\ell-iS\right)^L
  =\prod_{\alpha=1}^{N_\downarrow}
   {\lambda_\ell-\nu_\alpha'+{i\over2}\over
	 \lambda_\ell-\nu_\alpha'-{i\over2}}\ .
\label{app:bae11}
\end{equation}
We continue by rewriting these equations as $Q(\lambda_\ell)=0$ with
\begin{equation}
  Q(\omega) = (\omega+iS)^L \prod_{\alpha=1}^{N_\downarrow}
	\left(\omega - \nu_\alpha' -{i\over2}\right)
      - (\omega-iS)^L \prod_{\alpha=1}^{N_\downarrow}
	\left(\omega - \nu_\alpha' +{i\over2}\right)\ .
\label{app:Qdef}
\end{equation}
Similar as above we can identify the first $N_h+N_\downarrow$ roots of
this polynomial of degree $L+N_\downarrow$ with $\lambda_j$ and denote
the remaining $L-N_h\equiv N_e$ ones by $\lambda_j'$.  Proceeding as
in (\ref{app:contour}) we obtain
\begin{eqnarray}
  \sum_{k=1}^{N_h+N_\downarrow}
	\ln \left({\nu_\alpha'-\lambda_k+{i\over2}\over
	           \nu_\alpha'-\lambda_k-{i\over2}}\right)
 &=& \sum_{k=1}^{N_h+N_\downarrow} {1\over2\pi i}\oint_{C_k}{\rm d}z
        \ln \left({\nu_\alpha'-z+{i\over2}\over
	     \nu_\alpha'-z-{i\over2}}\right)
	{{\rm d}\over{\rm d}z} \ln Q(z)
\nonumber\\
 &=& -\sum_{k=1}^{N_e}
	\ln \left({\nu_\alpha'-\lambda_k'+{i\over2}\over
	           \nu_\alpha'-\lambda_k'-{i\over2}}\right)
	+ \ln\left(Q(\nu_\alpha'+{i\over2})/
		   Q(\nu_\alpha'-{i\over2})\right)\ .
\end{eqnarray}
Exponentiating this equation we obtain
\begin{equation}
  \prod_{k=1}^{N_h+N_\downarrow}
	{\nu_\alpha'-\lambda_k+{i\over2}\over
         \nu_\alpha'-\lambda_k-{i\over2}}
 =\prod_{k=1}^{N_e}
	{\nu_\alpha'-\lambda_k'-{i\over2}\over
	 \nu_\alpha'-\lambda_k'+{i\over2}}\,\,
	{Q(\nu_\alpha'+{i\over2})\over  Q(\nu_\alpha'-{i\over2})}\ .
\end{equation}
Using this and the fact that $\nu_\alpha'$ solve the second set of
Eqs.~(\ref{bae3}) together with the definition (\ref{app:Qdef}) the
unprimed variables can be eliminated and we find
\begin{equation}
   \left(\nu_\alpha'+i(S-{1\over2})\over 
	 \nu_\alpha'-i(S-{1\over2})\right)^L
 = -\prod_{\beta=1}^{N_\downarrow}
   {\nu_\alpha'-\nu_\beta'+i \over \nu_\alpha'-\nu_\beta'-i}
   \prod_{k=1}^{N_e}
   {\nu_\alpha'-\lambda_k'-{i\over2}\over
	 \nu_\alpha'-\lambda_k'+{i\over2}}\ .
\label{app:bae12}
\end{equation}
Comparing Eqs.~(\ref{app:bae11}) and (\ref{app:bae12}) with the Bethe
Ansatz equations (\ref{bae1}) the equivalence of the latter with
(\ref{bae3}) becomes evident.  The proof of equivalence with
(\ref{bae2}) is completely analogous.

\setlength{\baselineskip}{14pt}

\end {document}